\documentclass[english]{sbrt}
\usepackage[english]{babel}
\usepackage[utf8]{inputenc}

\usepackage{graphicx}
\usepackage[svgnames]{xcolor}
\usepackage{amsmath,amsfonts,amssymb,bm}
\usepackage[ruled,vlined]{algorithm2e}
\usepackage{algpseudocode}
\usepackage[nolist,nohyperlinks]{acronym} 
\usepackage{subfig}

\begin{document}
\begin{acronym}
	\acro{IRS}{Intelligent reflecting surface}
	\acro{RIS}{Reconfigurable intelligent surface}
    \acro{RIs}{reconfigurable intelligent surface}
	\acro{irs}{intelligent reflecting surface}
	\acro{PARAFAC}{parallel factor}
	\acro{TALS}{trilinear alternating least squares}
	\acro{BALS}{bilinear alternating least squares}
    \acro{ALS}{alternating least squares}
	\acro{DF}{decode-and-forward}
	\acro{AF}{amplify-and-forward}
	\acro{CE}{channel estimation}
	\acro{RF}{radio-frequency}
	\acro{THz}{Terahertz communication}
	\acro{EVD}{eigenvalue decomposition}
	\acro{CRB}{Cramér-Rao lower bound}
	\acro{CSI}{channel state information}
	\acro{BS}{base station}
	\acro{MIMO}{multiple-input multiple-output}
	\acro{NMSE}{normalized mean square error}
	\acro{2G}{Second Generation}
	\acro{3G}{3$^\text{rd}$~Generation}
	\acro{3GPP}{3$^\text{rd}$~Generation Partnership Project}
	\acro{4G}{4$^\text{th}$~Generation}
	\acro{5G}{5$^\text{th}$~Generation}
	\acro{6G}{6$^\text{th}$~generation}
	\acro{E-TALS}{\textit{enhanced} TALS}
	\acro{UT}{user terminal}
	\acro{UTs}{users terminal}
	\acro{LS}{least squares}
	\acro{KRF}{Khatri-Rao factorization}
	\acro{KF}{Kronecker factorization}
	\acro{MU-MIMO}{multi-user multiple-input multiple-output}
	\acro{MU-MISO}{multi-user multiple-input single-output}
	\acro{MU}{multi-user}
	\acro{SER}{symbol error rate}
	\acro{SNR}{signal-to-noise ratio}
	\acro{SVD}{singular value decomposition}
        \acro{DFT}{Discrete Fourier Transform}
\end{acronym}
\title{Circuit-Based Modeling Approach for Channel Estimation in RIS-Assisted Communications}

\author{Daniel C. Alcantara, Daniel V. C. de Oliveira, Gilderlan T. de Araújo\\ Paulo R. B. Gomes, André L. F. de Almeida
\thanks{Daniel C. Alcantara, Daniel V. C. de Oliveira, Gilderlan T. de Araújo and André L. F. de Almeida are with the Teleinformatics Department, Federal University of Ceará, Fortaleza-CE, e-mail: \{danielchaves,danielvctor\}@alu.ufc.br; \{gilderlan,andre\}@gtel.ufc.br}
\thanks{Paulo. R. B. Gomes is with the Federal Institute of Ceará, Tauá-CE, e-mail: gomes.paulo@ifce.edu.br.}
\thanks{The authors acknowledge the partial support of FUNCAP under grant ITR-0214-00041.01.00/23, and the National Institute of Science and Technology (INCT-Signals) sponsored by Brazil's National Council for Scientific and Technological Development (CNPq) under grant 406517/2022-3. This work is also partially supported by CNPq under grant 312491/2020-4 and by Ericsson Research, Sweden, and Ericsson Innovation Center, Brazil, under UFC.52 Technical Cooperation Contract Ericsson/UFC.}%
}

\maketitle


\markboth{XLIII BRAZILIAN SYMPOSIUM ON TELECOMMUNICATIONS AND SIGNAL PROCESSING - SBrT 2025, SEPTEMBER 29TH TO OCTOBER 2ND, NATAL, RN}{}

\begin{abstract}

\acl{RIS} (RIS) has been explored as a supportive technology for wireless communication since around 2019. While the literature highlights the potential of RIS in different modern applications, two key issues have gained significant attention from the research community: channel estimation and phase shift optimization. The performance gains of RIS-assisted systems rely heavily on optimal phase shifts, which, in turn, depend on accurate channel estimation. Several studies have addressed these challenges under different assumptions. Some works consider a range of continuous phase shifts,
while others propose a limited number of discrete phase values for the RIS elements. Many studies present an idealized perspective, whereas others aim to approximate more practical aspects by considering circuit system responses and employing phase shifts derived from a Discrete Fourier Transform (DFT) or other lookup tables. However, to our knowledge, no study has examined the influence of circuit system parameters on channel estimation and subsequent phase shift optimization. This paper models each RIS element as an equivalent resonant circuit composed of resistance, capacitance, and inductance. We propose that resistance and capacitance parameters can be dynamically and independently configured, leading to the formulation of an impedance matrix. Furthermore, we construct a circuit-based RIS phase shift matrix that accounts for the response of the resonant circuit, which changes with variations in the physical parameters of resistance and capacitance. We investigate the impact of this circuit-based RIS phase shift within a tensor-based channel estimation approach. Our results indicate a performance loss compared to ideal scenarios, such as those using the DFT design. However, we found that increasing the training time can mitigate this performance degradation.

\end{abstract}

\renewcommand\baselinestretch{.85}

\vspace{-0.3cm}
\section{Introduction}
The concept of \ac{RIS} for wireless communication emerged around 2019 \cite{Basar_2019}. The potential of RIS technology lies in its ability to reduce randomness in the wireless propagation environment by controlling in real-time the wave propagation characteristics, such as amplitude and phase \cite{Di_Renzo_2020}. Since its conception, different modern applications utilizing RIS as support technology have been developed. 

Among the main challenges facing RIS-assisted wireless communications, channel estimation is one of the most critical issues to be tackled due to the passive nature of the RIS. Accurate channel state information (CSI) is essential for optimizing both the passive (phase shift) and active beamformers to maximize the system's spectral efficiency during the data transmission phase. Many works have studied the channel estimation problem from different perspectives, such as different propagation scenarios and signal processing methods \cite{Lee_2022, Zhang_2022}.


The work \cite{on-off} introduced an early method for estimating the cascaded channel (element by element) between the user terminal-RIS-base station by toggling the individual RIS elements between ON and OFF states. In \cite{1Singh_2024}, the RIS elements were subdivided into groups, and a graph transformer-based method was proposed. This approach exploited the spatial correlation of groups in the channel estimation solution.
Additionally, \cite{Rapudu_2025} presented a machine learning method for estimating the uplink cascaded time-varying channel in a multi-RIS scenario. Focusing on a structured channel model, \cite{1Fazal_2024} exploited the low-rank tensor structure of the received signal to estimate the channel parameters in a decoupled manner across two spatial dimensions. In contrast, \cite{Gil_JTSP} introduced the PARAFAC tensor decomposition to estimate the unstructured channels through closed-form and iterative solutions separately. While the works mentioned above assume RIS operates perfectly,  it is reasonable to analyze the impact of hardware/environmental impairments in the RIS elements on the channel estimation accuracy. In this way, \cite{Paulo_2023} proposed tensor-based solutions for RIS operating under different imperfections. In contrast to previous works, to improve spectral efficiency, \cite{Huang_2022} and \cite{Gil_TSP} derived semi-blind receivers.

Channel estimation is crucial to the phase shift optimization (commonly called passive beamforming optimization in the literature). Therefore, many works propose different solutions for this task.
For instance, \cite{jensen2019optimal} has demonstrated that the optimal phase shifts can be modeled as a Discrete Fourier Transform (DFT) matrix. Additionally, \cite{Yuri} developed a tensor-based framework for phase shift optimization that accounts for imperfect channel estimation. In both works, continuous phase shifts are assumed. In contrast, some works present optimization processes based on discrete phase shifts to address more practical applicability. Notable works in this area include \cite{Zhang_2022,Donglai_2024,Lucas}. Particularly, in \cite{Zhang_2022}, individual RIS elements are designed as an equivalent resonant circuit. In \cite{Lucas}, a set of discrete phases is approximated by DFT points, whereas \cite{Zhang_2022} proposes an analytical expression for designing phases, where amplitude and phase are coupled.


It is essential to mention that no previous work has addressed a phase shift design purely based on the actual circuit model. In contrast, in this paper, we propose a novel phase shift design approach that leverages the effects of variations in resistance and capacitance parameters for each RIS element. The contributions of this work are outlined as follows:

\begin{itemize}
     \item We propose a circuit-based phase shift matrix that considers the effects of resistance and capacitance variations. First, we select two vectors representing the resistance and capacitance. We calculate an impedance matrix from these vectors, which is then mapped to a phase shift matrix. The variation range for resistance and capacitance values is chosen to minimize the dissipation loss while maximizing the available phase range for configuring the RIS elements.

    \item We examine how the proposed circuit-based phase shift design affects the channel estimation accuracy. Specifically, we use a tensor-based channel estimation method that employs a PARAFAC-based receiver. 
\end{itemize}

\vspace{2ex}
{\color{black}\textit{Notation}: The font types $a$, $\mathbf{a}$, $\mathbf{A}$ and $\mathcal{A}$ denote scalar, vector, matrix, and tensor, respectively.
Transpose, pseudo-inverse and Frobenius norm of $\mathbf{A} \in \mathbb{C}^{I \times J}$ are represented by $\mathbf{A}^{\text{T}}$, $\mathbf{A}^\dagger$ and $\|\cdot\|_{\text{F}}$, respectively. The operator $\text{diag}(\mathbf{a})$ forms a diagonal matrix out of its vector argument. The symbol $\diamond$ represents the Khatri-Rao product. $\mathbf{A}_{ij}$, $\mathbf{A}_{i\cdot}$ and $\mathbf{A}_{\cdot j}$ stand for the $(i,j)$-th element, $i$-th row and the $j$-th column of $\mathbf{A}$, respectively. $\mathcal{I}_{3,R}$ is the third-order identity tensor of size $R \times R \times R$, while $\times_n$ is the $n$-mode product between a tensor and a matrix.}

\section{System Model}
Let us consider the uplink of a narrowband RIS-assisted MIMO system with \(M_t\) transmitter antennas, \(M_r\) receiver antennas, and \(N\) RIS elements. The link between the user terminal (UT) and base station (BS) is considered to be either unavailable or negligible.
The transmission protocol follows
\cite{Gil_JTSP}. Specifically, we assume a block transmission where the total channel estimation time \(T_E\) is divided into \(K\) blocks, each consisting of \(T\) time slots (\(T_E = KT\)).
By assuming a RIS with varying phase shifts from block to block, the received signal during block \(k\) at time slot \(t\) is expressed as
\begin{equation}
    \mathbf{y}_{k,t} = \mathbf{H}\mathbf{S}_k\mathbf{G}^{\text{T}}\mathbf{x}_{t} + \mathbf{z}_{k,t},
\end{equation}
such that after $T$ pilot transmissions the received signal at block $k$ is given as
\begin{equation}
    \mathbf{Y}_{k} = \mathbf{H}\text{diag}(\mathbf{s}_k)\mathbf{G}^{\text{T}}\mathbf{X} + \mathbf{Z}_k,
\end{equation}
where \(\mathbf{H} \in \mathbb{C}^{M_r \times N}\) represents the channel between the BS and the \ac{RIS}, and \(\mathbf{G} \in \mathbb{C}^{M_t \times N}\) denotes the channel between the UT and the \ac{RIS}, respectively. The matrix \(\mathbf{X} = [\mathbf{x}_1 \; \mathbf{x}_2 \; \dots \; \mathbf{x}_T] \in \mathbb{C}^{M_t \times T}\) collects the transmitted pilots. Additionally, the phase shift vector applied at the RIS during block \(k\) is represented as \(\mathbf{s}_k = [s_{1} e^{j\theta_1}, \ldots, s_{N} e^{j\theta_N}]^{\mathrm{T}} \in \mathbb{C}^{N \times 1}\). The phase shift matrix which contains all the $K$ phase shift patterns used during the channel estimation task is represented by \(\mathbf{S} = [\mathbf{s}_1 \; \dots \; \mathbf{s}_K]^{\text{T}} \in \mathbb{C}^{K \times N}\). The matrix $\mathbf{Z}_{k}$ represents the additive white Gaussian noise (AWGN). Given that the pilot matrix \(\mathbf{X}\) is semi-unitary, we can filter the received signal at block \(k\) as follows
\begin{equation}
    \overline{\mathbf{Y}}_{k} = \mathbf{H}\text{diag}(\mathbf{s}_k)\mathbf{G}^{\text{T}}\ + \mathbf{Z}_k\mathbf{X}^{\text{T}},
    \label{EQ: received signal1}
\end{equation}
where $\mathbf{Z}_k\mathbf{X}^{\text{T}}$ denotes the filtered noise term\footnote{From this point, we will disregard the noise term for clarity. However, it is important to note that we use the noisy version of the received signal in our computer simulations.}. According to \cite{Kolda_2009}, the Equation 
(3) represents the $k$-th frontal slice\footnote{In this paper, the definitions and operations involving tensors follows \cite{Kolda_2009}.} of the received signal tensor $\overline{\mathcal{Y}} \in \mathbb{C}^{M_r \times M_t \times K}$, which follows a rank-$N$ third-order PARAFAC model, i.e., 
\begin{equation}
\overline{\mathcal{Y}} = \mathcal{I}_{3,N} \times_1 \mathbf{H} \times_2 \mathbf{G} \times_3 \mathbf{S} \in \mathbb{C}^{M_r \times M_t \times K}.
\label{EQ:n-mode notation}
\end{equation}

\vspace{-0.5cm}
\section{RIS Phase Shift Matrix Design}
\begin{figure}
    \centering
    \includegraphics[width=0.75\linewidth]{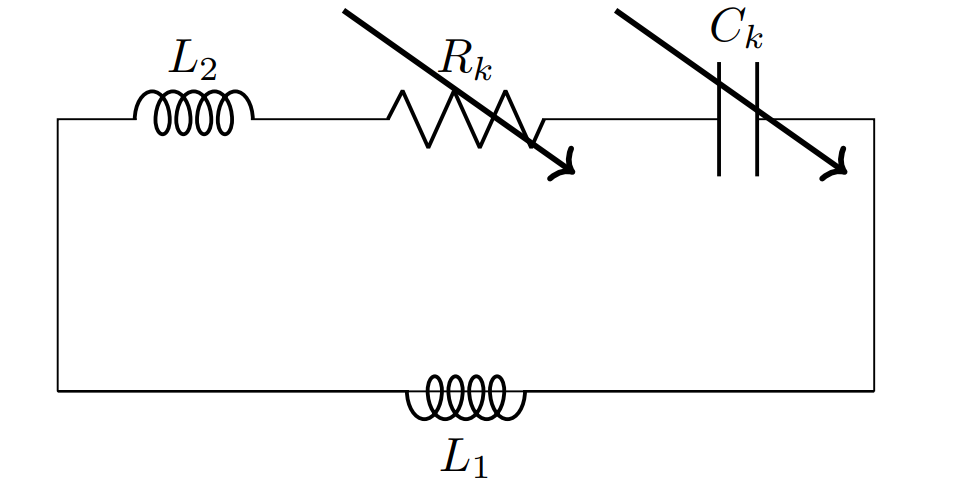}
    \caption{Varactor diode equivalent circuit model.}
    \label{fig:designhardware}
\end{figure}

Most channel estimation methods for RIS-assisted wireless communications assume an ideal design for the phase shift matrix. According to \cite{jensen2019optimal}, the DFT matrix is an optimal design for channel estimation (for more details, access the reference). In a less restricted way, other works restrict only the amplitude response, considering unitary amplitude (i.e., perfect full signal reflection by each RIS element).
However, achieving both DFT design and unitary amplitude responses is challenging in practice due to limitations in real-world circuit components in which the amplitude variation is phase-dependent. Aiming for a more realistic approach, \cite{Rui_Zhand_Practical_TCOM} proposes a physical modeling where each RIS element is designed as an equivalent parallel resonant circuit (illustrated in Figure \ref{fig:designhardware}) with the following impedance characteristics

\begin{equation}\label{EQ: impedancia}
    Z_n(C_n, R_n) = \frac{j\omega L_1 \left(j\omega L_2 + \frac{1}{j\omega C_n} + R_n\right)}{j\omega L_1 + \left(j\omega L_2 + \frac{1}{j\omega C_n} + R_n\right)}\;,
\end{equation}
where \(L_1\) and \(L_2\) represent the inductances, while \(C_n\) and \(R_n\) denote the effective capacitance and resistance of the $n$-th RIS element, respectively, and \(\omega\) denotes the angular frequency of the incident electromagnetic wave. From (\ref{EQ: impedancia}), we can note that 
full reflection is practically impossible due to dissipation losses in the circuit components (which cannot be zero in practice). Being \(Z_0\) the free-space impedance, 
the reflection coefficient of the $n$-th RIS element can be expressed as follows
\begin{equation}\label{EQ:coefReflec1}
v_n = \frac{Z_n(C_n,R_n) - Z_0}{Z_n(C_n,R_n) + Z_0}\;.
\end{equation}

Thus, the amplitude and phase shift response of the $n$-th RIS element are obtained as  
\( |v_n| \) and \( \arg(v_n) \), respectively. Looking at (\ref{EQ:coefReflec1}), it is possible to observe that
the phase and amplitude responses are interconnected and depend on the choices of \( R_n \) and \( C_n \), which both can be controlled independently and dynamically \cite{liu2019}. To illustrate the implications of this practical phase shift design, let us consider the following example parameters from \cite{Rui_Zhand_Practical_TCOM}: \( L_1 = 2.5 \, \text{nH} \), \( L_2 = 0.7 \, \text{nH} \), \( Z_0 = 377 \, \Omega \), and \( f = 2.4 \times 10^{9} \, \text{Hz} \). The considered capacitance range is between \( 0.47 \, \text{pF} \) and \( 2.35 \, \text{pF} \), while the resistance range varies from \( 0.5 \, \Omega \) to \( 2.5 \, \Omega \).
\begin{figure}
    \centering
    \includegraphics[width=1.1\linewidth]{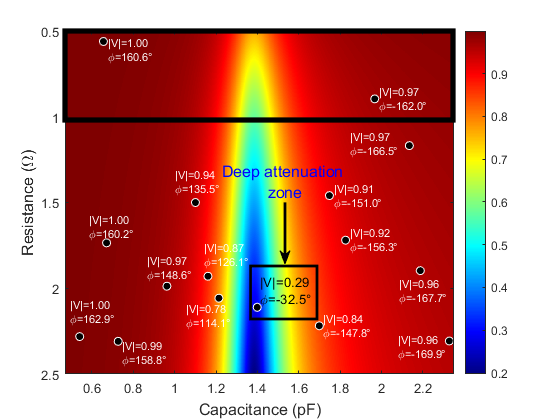}
    \caption{Amplitude Heatmap - Resistance versus Capacitance}
    \label{fig:mapadecaloramplitude}
\end{figure}
\begin{figure}
    \centering
    \includegraphics[width=1.1\linewidth]{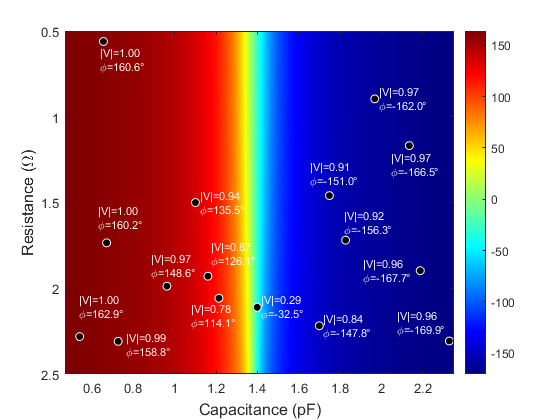}
    \caption{Phase Heatmap - Resistance versus Capacitance}
    \label{fig:mapadecalorphase}
\end{figure}
The heatmaps illustrated in Figures \ref{fig:mapadecaloramplitude} and \ref{fig:mapadecalorphase} enable us to analyze jointly the amplitude and phase responses of the RIS elements based on their component measurements. Figure \ref{fig:mapadecaloramplitude} illustrates a region of significant dissipation, which characterizes a strong attenuation zone. To demonstrate this, we consider $N = 16$ RIS elements (black points) in which the values of \( R_n \) and \( C_n \) were randomly generated within the considered ranges. 
{\color{black}In this figure, we identify a point located in the attenuation zone and another nearby.} 
By comparing Figures \ref{fig:mapadecaloramplitude} and \ref{fig:mapadecalorphase}, we can see that lower values of \( C_n \) correspond to a region where the amplitude response is close to 1. However, in this same region, the phase range is quite restricted. This practical limitation could pose a challenge for the RIS operation, since its fundamental purpose is manipulating phases to perform passive beamforming effectively.

\subsection{Proposed Phase Shift Design}
\label{sec:Proposed S}
Based on the previous discussion, we can identify an optimal operating region for the RIS by analyzing the heatmap results. Our goal is to select an RIS work region that jointly maximizes the phase range while minimizing attenuation. This represents an inherent trade-off to consider in the practical model. In our study, we have selected the highlighted area (black box) in Figure \ref{fig:mapadecaloramplitude} to bypass the zone of significant dissipation, specifically when \( R_n \) falls within the \( [0.5 \, \Omega, 1 \, \Omega] \) range and \( C_n \) is in the range of \( [1 \, \text{pF}, 2 \, \text{pF}] \). Within this interval, we define the phase vector \( \mathbf{c_n} \in \mathbb{C}^{1 \times N} \) and the resistance vector \( \mathbf{r_n} \in \mathbb{C}^{1 \times N} \), where \( N \) points are uniformly distributed.
By applying a circular shift, we can construct the resistance matrix $\mathbf{R} \in \mathbb{C}^{K \times N}$ and the capacitance matrix $\mathbf{C} \in \mathbb{C}^{K \times N}$ as follows
\begin{equation}
    \mathbf{R}_{k.} = \text{circshift}(\mathbf{r_n}, N) \quad \text{and} \quad \mathbf{C}_{k.} = \text{circshift}(\mathbf{c_n}, N).
    \label{EQ:R matrix}
\end{equation}
From $\mathbf{R}$ and $\mathbf{C}$, we can obtain 
the impedance matrix $\mathbf{Z} \in \mathbb{C}^{K \times N}$ using (\ref{EQ: impedancia}). The $(k,n)$-th element of $\mathbf{Z}$ is given by 
\begin{equation}\label{EQ: impedancia_applyied}
    \mathbf{Z}_{kn}(\mathbf{C}, \mathbf{R}) = \frac{j\omega L_1 \left(j\omega L_2 + \frac{1}{j\omega \mathbf{C}_{kn}} + \mathbf{R}_{kn}\right)}{j\omega L_1 + \left(j\omega L_2 + \frac{1}{j\omega \mathbf{C}_{kn}} + \mathbf{R}_{kn}\right)}\; .
\end{equation}
Let us define the proposed phase shift matrix $\mathbf{S} \in \mathbb{C}^{K \times N}$ whose its $(k,n)$-th element is given by


\begin{equation}
    \mathbf{S}_{kn} = A_{kn} \cdot e^{j\phi_{kn}},
    \label{EQ: RIS element}
\end{equation}
where $A_{kn} = |v_{kn}|$, $\phi = \arg(v_{kn})$ and



\begin{equation}\label{EQ:coefReflec}
v_{kn} = \frac{\mathbf{Z}_{kn}(\mathbf{C}, \mathbf{R}) - Z_0}{\mathbf{Z}_{kn}(\mathbf{C}, \mathbf{R}) + Z_0}\;.
\end{equation}

Note that our proposed phase shift design is derived from the response of the practical hardware model
without any reliance on analytical methods or DFT approximations. This raises an important question: How does this practical design impact channel estimation accuracy? In the next section, we briefly present the PARAFAC-based channel estimator that will use our phase shift design. In Section \ref{results}, the channel estimation accuracy is evaluated through computational experiments by considering practical phase shift responses.



\section{Channel Estimation Approach}
The proposed phase shift design, discussed in Section \ref{sec:Proposed S}, can be applied to any RIS channel estimation approach. Here, we capitalize on the tensor structure of the received signal in (\ref{EQ: received signal1})\footnote{\textcolor{black}{Tensor signal processing has successfully been applied to solve different problems in the wireless communication area such as cooperative relay systems \cite{Ximenes_TSP_2014}, tensor coding \cite{Favier_TSP_2014} and multiuser equalization \cite{DEALMEIDA_2007}, to mention a few.}}. With this strategy, we will evaluate the impact of our proposed phase shift design on channel estimation accuracy using the pilot-assisted PARAFAC-based receiver formulated by \cite{Gil_SAM,Gil_JTSP}, which is summarized below.



The estimation of the involved channels is obtained by minimizing the following objective function:
\begin{equation}
   (\hat{\mathbf{H}}\; ,\hat{\mathbf{G}}) = \underset{\mathbf{H},\mathbf{G}}{\arg\min} \,\, \left\|\overline{\mathbf{Y}}_{k} - \mathbf{H}\text{diag}(\mathbf{s}_k)\mathbf{G}^{\textrm{T}}\right\|_\text{F}^2,
   \label{EQ: Global OPT problem} 
\end{equation}
which represents a nonlinear problem. To solve (\ref{EQ: Global OPT problem}), we exploit the unfolding representations of $\overline{\mathcal{Y}}$. Specifically, the $1$-mode and $2$-mode unfoldings that are given by
\begin{equation}\label{modo1}
[\overline{\mathcal{Y}}]_{(1)} = \mathbf{H} \left( \mathbf{S} \diamond \mathbf{G} \right)^{\text{T}} \quad \text{and} \quad [\overline{\mathcal{Y}}]_{(2)} = \mathbf{G} \left( \mathbf{S} \diamond \mathbf{H} \right)^{\text{T}}.
\end{equation}
From (\ref{modo1}), the estimates $\hat{\mathbf{H}}$ and $\hat{\mathbf{G}}$ can be obtained 
using the well known \ac{ALS} algorithm \cite{comon_2009}). It consists of solving the following sub-optimal LS problems in an iterative and alternating manner
\begin{eqnarray}
&\hat{\mathbf{H}} = \underset{\mathbf{G}}{\arg\min} \,\, \left\|[\overline{\mathcal{Y}}]_{(1)} - \mathbf{H}(\mathbf{S} \diamond \mathbf{G})^{\text{T}}\right\|_{\text{F}}^2\label{func costG},\\
&\hat{\mathbf{G}} = \underset{\mathbf{H}}{\arg\min}\,\, \left\|[\overline{\mathcal{Y}}]_{(2)} - \mathbf{G}(\mathbf{S} \diamond \mathbf{H})^{\text{T}}\right\|_\text{F}^2.
\label{Func costZ}
\end{eqnarray}
The solutions of (\ref{func costG}) and (\ref{Func costZ}) are given by
\begin{equation}
    \hat{\mathbf{H}} = [\overline{\mathcal{Y}}]_{(1)}[(\mathbf{S} \diamond \mathbf{G})^{\text{T}}]^\dag \quad \text{and} \quad \hat{\mathbf{G}} = [\overline{\mathcal{Y}}]_{(2)}[(\mathbf{S} \diamond \mathbf{H})^{\text{T}}]^\dag,
\end{equation}
respectively. To better understand, we present in Algorithm \ref{alg:bals} the pseudo-code of the PARAFAC-based channel estimation procedure.

\begin{algorithm}
\scriptsize{
\DontPrintSemicolon
\SetAlgoLined
\KwIn{\(i = 0\); initialize \(\widehat{\mathbf{G}}\)}
\KwOut{Estimated matrices \(\widehat{\mathbf{H}},\widehat{\mathbf{G}}\)}
\For{\(i = 1\) \KwTo \texttt{iter}}{
      \(\widehat{\mathbf{H}}^{(i)} \gets [\overline{\mathcal{Y}}]_{(1)}\left(\left(\mathbf{S} \diamond \widehat{\mathbf{G}}^{(i)}\right)^\text{T}\right)^{\dagger}\)\;
    
    \(\widehat{\mathbf{G}}^{(i)} \gets [\overline{\mathcal{Y}}]_{(2)}\left(\left(\mathbf{S} \diamond \widehat{\mathbf{H}}^{(i-1)}\right)^\text{T}\right)^{\dagger}\)\;
        
    Compute the relative reconstruction error:\;
    
    \(e(i) = \frac{\|[\overline{\mathcal{Y}}]_{(1)} - \widehat{\mathbf{H}}^{(i)}(\mathbf{S} \diamond \widehat{\mathbf{G}}^{(i)})^{\text{T}}\|_\text{F}^2}{\|[\overline{\mathcal{Y}}]_{(1)}\|_\text{F}^2}\)\;
    
    \If{\(i > 1\) \textbf{and} \(|e(i) - e(i-1)| \leq \varepsilon\)}{
        \textbf{break}\;
    }
}

\caption{PARAFAC-based channel estimator}
\label{alg:bals}
}
\end{algorithm}

The complexity and uniqueness issues of this tensor-based channel estimator are detailed in \cite{Gil_JTSP}. In the next section, we present simulation results to evaluate the channel estimation performance considering the proposed phase shift design in Section \ref{sec:Proposed S} and Algorithm \ref{alg:bals}. 


\begin{figure}[!t]
    \centering
    \subfloat[\vspace{-1ex} Amplitude distribution]{\includegraphics[scale=0.5]{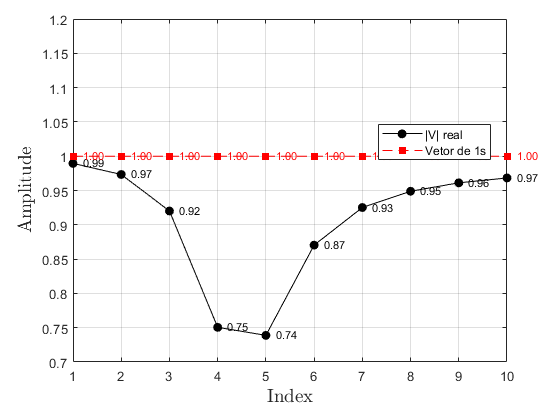}\label{Fig: NMSE}} \\
    \subfloat[\vspace{-1ex} Phases distribution]{\includegraphics[scale=0.5]{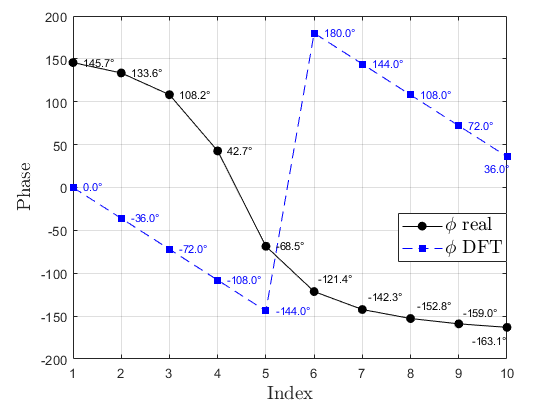}\label{Fig: complexity}}
    \caption{Example of amplitude and phase distributions considering uniform choice.}
    \label{Fig: example uniform}
\end{figure}

\begin{figure}[!t]
    \centering
    \subfloat[\vspace{-1ex} NMSE of $\hat{\mathbf{H}}$ vs. SNR (dB).]{\includegraphics[scale=0.49]{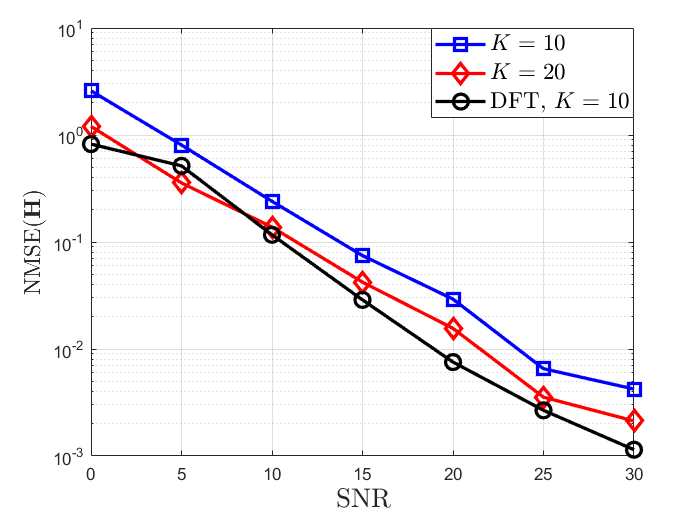}\label{fig: NMSE H}} \\
    \subfloat[\vspace{-1ex} NMSE of $\hat{\mathbf{G}}$ vs. SNR (dB).]{\includegraphics[scale=0.49]{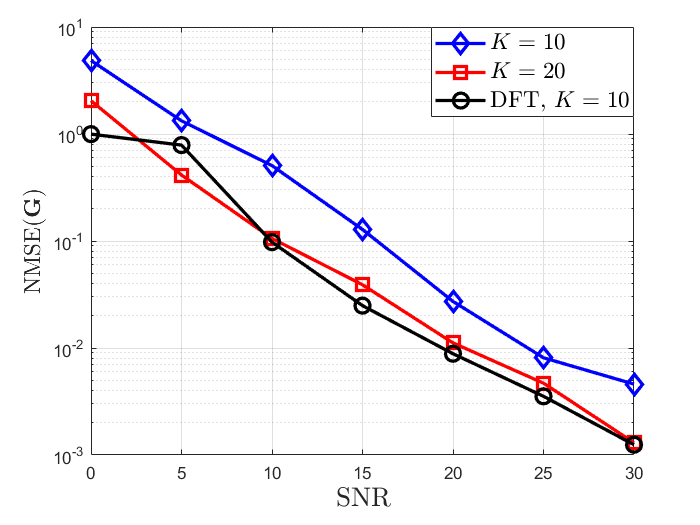} \label{fig:NMSEG}}
    \caption{System performance considering the circuit-based phase shift design.}
    \label{Fig: distribution}
\end{figure}
\section{Simulation Results}\label{results}
This section discusses the impact of the circuit-based phase shift design using the PARAFAC-based channel estimator in \cite{Gil_JTSP}. The performance is evaluated through the normalized mean square error (NMSE) of the estimated channels $\hat{\mathbf{H}}$ and $\hat{\mathbf{G}}$. The simulation setup consists of \( M_t = M_r = N = T = 10 \) and \( K \in \{10, 20\} \) and Rayleigh channels are assumed. The capacitance and resistance values are uniformly distributed within the ranges \([1 \text{ pF}, 2 \text{ pF}] \) and \([0.5 \, \Omega, 1 \, \Omega] \), as illustrated in Figure \ref{Fig: example uniform}. These range values are chosen to avoid deep energy dissipation (i.e., the amplitude response is close to 1, which implies stronger reflection) and to maximize the available phase for RIS elements configuration during the channel estimation task.

Figures \ref{fig: NMSE H} and \ref{fig:NMSEG} show the channel estimation accuracy for the involved channels obtained from Algorithm \ref{alg:bals}. It is observed that, in both results, the estimation performance decreases when the practical circuit-based phase shift model (represented by blue and red curves) is considered compared to RIS elements configured by DFT matrix (black curves) during the $K$ blocks. This result is expected since for DFT design, each reflecting element has zero energy dissipation, and the correlation properties of the additive noise are not affected by filtering at the receiver performed in Equation (\ref{EQ: received signal1}). However, in practice, this is an idealized assumption because semiconductor devices, dielectrics, and metals used in manufacturing the RIS elements dissipate some energy level. On the other hand, the performance gap between practical circuit-based and DFT phase shift designs decreases when the training time increases, showing that the circuit-based phase shift design can perform very close to the DFT one. Therefore, a trade-off can be observed, revealing that increasing the training time can help mitigate the losses in the channel estimation performance in practical scenarios.



\section{Conclusion}
We proposed a new design for a phase shift matrix that incorporates a circuit-based model, considering each RIS element's circuit response. We examine how the proposed phase shift design affects the channel estimation accuracy of a tensor-based receiver. Our results indicate a performance loss of the practical circuit model compared to the idealized DFT design. However, we observed that this loss can be minimized by increasing the duration of the training. This presents a trade-off, as achieving the highest channel estimation accuracy requires optimal phase shift settings in the practical RIS. Looking ahead, we plan to investigate the best operational region for this design, explore how to select the resistance and capacitance vectors effectively, and assess the impact of this design on other types of receivers, including semi-blind receivers that enhance spectral efficiency.


\bibliographystyle{IEEEtran}
\bibliography{IEEEexample}

\end{document}